\documentclass[10pt,conference]{IEEEtran}
\IEEEoverridecommandlockouts

\usepackage[most]{tcolorbox}
\usepackage{amsmath,graphicx,tikz,pifont,fontawesome5}  \usepackage[caption=false,font=footnotesize]{subfig}
\usetikzlibrary{shapes.multipart}
\usetikzlibrary{positioning,arrows.meta,fit,backgrounds,calc}
\usepackage{cite}
\usepackage{amsmath,amssymb,amsfonts}
\usepackage{algorithmic}
\usepackage{graphicx}
\usepackage{textcomp}
\usepackage{xcolor}
\usepackage{fancyvrb}
\usepackage{xspace}
\usepackage[hidelinks]{hyperref}
\usepackage{booktabs}
\usepackage{graphicx}
\usepackage{booktabs}
\usepackage{xspace}
\usepackage[most]{tcolorbox}
\usepackage{amsmath,graphicx,tikz,pifont,fontawesome5}  \usepackage[caption=false,font=footnotesize]{subfig}
\usetikzlibrary{shapes.multipart}
\usetikzlibrary{positioning,arrows.meta,fit,backgrounds,calc}

\def\BibTeX{{\rm B\kern-.05em{\sc i\kern-.025em b}\kern-.08em
    T\kern-.1667em\lower.7ex\hbox{E}\kern-.125emX}}

\newcommand{\dfuzz}{\textsc{Difuzzi}\xspace}
\newcommand{\ignore}[1]{}

\begin{document}

\bstctlcite{IEEEexample:BSTcontrol}

\title{\dfuzz: A Differential Fuzzing Tool for Code Functional Similarity Assessment}

\author{\IEEEauthorblockN{Simantika Bhattacharjee Dristi}
\IEEEauthorblockA{\textit{University of Virginia} \\
Charlottesville, Virginia, USA \\
nwc8gr@virginia.edu}
\and
\IEEEauthorblockN{Matthew B. Dwyer}
\IEEEauthorblockA{\textit{University of Virginia} \\
Charlottesville, Virginia, USA \\
matthewbdwyer@virginia.edu}
}

\maketitle

\begin{abstract}
A range of software engineering tasks, such as optimization and refactoring, seek to preserve the behavior of software components and require methods to assess functional similarity. Research has explored formal approaches to \textit{prove} functional equivalence and used machine learning methods
to \textit{approximate} it, but the former does not scale and does not quantify the extent of similarity and the latter struggles when subtle semantic differences are present.  Differential fuzzing offers a scalable approach for differencing programs semantically, but applying it often requires substantial manual effort to construct input generators that are well-suited to a target pair of programs. To address this challenge, we introduce \dfuzz, which incorporates an LLM-based generator synthesis step that is iteratively validated and repaired.
By automating this step, \dfuzz paves the way for broader applicability of differential fuzzing. We release \dfuzz at \url{https://github.com/Dristi123/difuzzi_tool}, with a demonstration video at \url{https://youtu.be/FlHNOHuo9WE}
\end{abstract}

\section{Introduction}
Assessing the functional correctness of candidate programs against reference implementations is fundamental to a wide range of software engineering (SE) tasks such as code generation, refactoring, optimization, and translation~\cite{diffuzz_refac, 10795375, shypula2024learning}. While binary correctness metrics distinguish correct from incorrect programs, continuous functional similarity scores capture the extent of behavioral agreement, providing finer-grained feedback for debugging and iterative code improvement~\cite{codescore,debug}. However, existing approaches to functional similarity assessment have several limitations. Unit testing provides execution-based evidence but requires adequate predefined test suites. On the other hand, static code evaluation metrics (CEMs) like CodeBLEU~\cite{ren2020codebleumethodautomaticevaluation} offer inexpensive, execution-free evaluation but remain sensitive to surface-level similarities that may not reflect functional behavior~\cite{Local}. Similarly, LLM-judges can estimate similarity without execution~\cite{meeus2026detectingfunctionalmemorizationcode} but often struggle to reason about deeper program semantics~\cite{equibench}. Differential fuzzing offers a promising solution to these limitations by dynamically generating test inputs without requiring predefined test suites, while providing execution-based evidence of functional similarity that LLM-judges or static metrics lack. 
\par Despite its potential, differential fuzzing remains underexplored for code similarity and correctness evaluation. The few existing approaches~\cite{diffuzz_refac, Local} often require manual intervention, making large-scale analysis impractical, while the lack of open-source end-to-end tools further limits practical adoption.
\par To address these gaps, in this paper, we present \dfuzz, an automated, coverage-guided differential fuzzing tool that estimates functional similarity between source code implementations. Given a reference and a candidate , \dfuzz generates test inputs guided by coverage feedback from both implementations, executes them on the same inputs, and compares their outputs to assess functional agreement. Supporting Python and Java at both function and file levels, \dfuzz provides fine-grained functional similarity scores and also identifies concrete counterexamples that establish non-equivalence.

\par \dfuzz offers a fully automated workflow comprising LLM-based input generator synthesis, automated validation, and coverage-guided differential fuzzing. First, an LLM synthesizes a problem-specific generator that transforms raw fuzzer bytes into valid program inputs. \dfuzz then validates the LLM-synthesized generator for structural correctness, robustness, and input validity, using feedback-driven LLM repair to address failures. Once validated, the generator is integrated with a coverage-guided fuzzer, which mutates input bytes using coverage feedback from both implementations to exercise different execution paths and assess their functional agreement.

\par Our experiments in Section~\ref{sec:evaluation}, conducted on 500 Python and Java program pairs from 5 benchmarks, demonstrate \dfuzz's effectiveness in functional similarity estimation relative to baselines.

\section{Related Work}
Existing tools for comparing program behavior primarily focus on finding inputs that expose functional differences. For example, CrossHair~\cite{crosshair} employs differential symbolic execution to detect such differences, whereas Mokav~\cite{mokav} leverages LLMs and execution feedback to generate difference-exposing tests (DETs). While these approaches can identify counterexamples that establish non-equivalence, they do not indicate the extent of behavioral agreement between the two programs. 
\par The use of differential fuzzing for functional correctness assessment is not new. However, prior approaches ~\cite{Local, diffuzz_refac} are limited to Python, rely on manually validated input generators, and lack a publicly available, general-purpose tool. Furthermore, neither systematically evaluates the effectiveness of differential fuzzing itself for functional similarity estimation. This work addresses these gaps by introducing \dfuzz, which automates the workflow and supports both Python and Java, and by empirically evaluating \dfuzz's effectiveness against existing approaches.

\section{Framework}
\label{sec:framework}

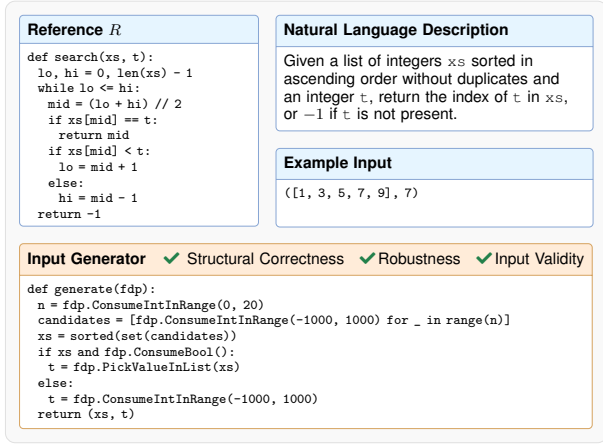
\begin{figure}[t]
\centering
\providecommand{\figtt}{\fontfamily{lmtt}\selectfont}
\providecommand{\figcode}[1]{{\figtt\scriptsize #1\par}}
\definecolor{userline}{RGB}{136,166,212}
\definecolor{llmline}{RGB}{214,163,96}%
\definecolor{okgreen}{RGB}{34,128,74}
\tikzset{
  card/.style={rectangle split, rectangle split parts=2, rounded corners=2pt, align=left, inner sep=4pt,
               font=\sffamily\footnotesize},
  user/.style={card, draw=userline, rectangle split part fill={blue!40!cyan!10, white}},
  llm/.style={card, draw=llmline, rectangle split part fill={orange!15, white}},
  panel/.style={draw=black!12, fill=black!2.5, rounded corners=5pt, inner sep=7pt},
}
\resizebox{0.9\columnwidth}{!}{%
\begin{tikzpicture}
  \node[user, text width=40mm, anchor=north west] (ref) at (0,0)
       {\textbf{Reference $R$}\nodepart{two}\figcode{%
def search(xs, t):\\
\ \ lo, hi = 0, len(xs) - 1\\
\ \ while lo <= hi:\\
\ \ \ \ mid = (lo + hi) // 2\\
\ \ \ \ if xs[mid] == t:\\
\ \ \ \ \ \ return mid\\
\ \ \ \ if xs[mid] < t:\\
\ \ \ \ \ \ lo = mid + 1\\
\ \ \ \ else:\\
\ \ \ \ \ \ hi = mid - 1\\
\ \ return -1}};
  \node[user, text width=54mm, anchor=north west] (desc) at ($(ref.north east)+(3mm,0)$)
       {\textbf{Natural Language Description}\nodepart{two} Given a list of integers \texttt{xs} sorted in ascending order without duplicates and an integer \texttt{t}, return the index of \texttt{t} in \texttt{xs}, or $-1$ if \texttt{t} is not present.};
  \path let \p1 = ($(desc.south)+(0,-3mm)$), \p2 = (ref.south), \p3 = (desc.north), \p4 = (desc.text split),
            \n1 = {\y1 - \y2 - (\y3 - \y4) - 8pt} in
    node[user, text width=54mm, anchor=north west] (exi) at ($(desc.south west)+(0,-3mm)$)
       {\textbf{Example Input}\nodepart{two}\parbox[t][\n1][t]{54mm}{\figcode{([1, 3, 5, 7, 9], 7)}}};

  \node[llm, text width=99.8mm, anchor=north west] (gen) at ($(ref.south west)+(0,-3mm)$)
       {\textbf{Input Generator}\hfill{\footnotesize{\color{okgreen}\faCheck}\, Structural Correctness\quad
        {\color{okgreen}\faCheck}\,Robustness\quad{\color{okgreen}\faCheck}\,Input Validity}
        \nodepart{two}\figcode{%
def generate(fdp):\\
\ \ n = fdp.ConsumeIntInRange(0, 20)\\
\ \ candidates = [fdp.ConsumeIntInRange(-1000, 1000) for \char95{} in range(n)]\\
\ \ xs = sorted(set(candidates))\\
\ \ if xs and fdp.ConsumeBool():\\
\ \ \ \ t = fdp.PickValueInList(xs)\\
\ \ else:\\
\ \ \ \ t = fdp.ConsumeIntInRange(-1000, 1000)\\
\ \ return (xs, t)}};
  \begin{scope}[on background layer]
    \node[panel, fit=(ref)(desc)(exi)(gen)] {};
  \end{scope}
\end{tikzpicture}}
\caption{Example of input generator synthesis (Section~\ref{sec: iput_generator}) }
\label{fig:example_step1}
\end{figure}

Given a reference program $R$, one or more candidate implementations of the same problem, at least one example input, and an optional NL description of $R$, \dfuzz estimates the functional similarity of each candidate $C$ to $R$. 

\textbf{Assumptions.} \textsc{Difuzzi} makes two assumptions about $R$ and $C$: (a) Each program is self-contained, i.e., includes all necessary dependencies (e.g., helper functions and imports), and can be executed in isolation; and (b) Each program produces a comparable output, through return values (function level) or standard output (file level). These outputs define the scope of observable behavior in the current version of \dfuzz.

\subsection{Approach}
\label{sec:approach}

\begin{figure}[t]
\centering
\resizebox{\linewidth}{!}{%
\begin{tikzpicture}[
    node distance=5.5mm,
    font=\sffamily\footnotesize,
    >={Stealth[length=1.8mm]},
    box/.style={
      draw, rounded corners=2pt, fill=white,
      align=center, minimum height=8mm, inner sep=2.5pt
    },
    inp/.style={box, fill=gray!8, text width=20mm},
    chk/.style={box, text width=21mm, minimum height=7mm},
    prog/.style={box, text width=17mm, fill=blue!6},
    stage/.style={
      draw, dashed, rounded corners=4pt, inner sep=5pt
    },
    title/.style={
      font=\sffamily\footnotesize\bfseries,
      fill=white, inner sep=2pt
    },
    flow/.style={->, semithick},
    note/.style={font=\sffamily\scriptsize, align=center},
  ]

  \node[inp] (exi) {Example inputs};
  \node[inp, above=3mm of exi] (ref) {Reference $R$};
  \node[inp, below=3mm of exi] (desc)
    {Description\\[-1pt]{\scriptsize(optional)}};

  \node[box, text width=10mm, fill=orange!12,
        right=9mm of exi] (llm)
    {{\Large\faRobot}\\[1pt]LLM};

  \node[box, text width=27mm, right=7mm of llm] (gen)
    {Input generator};

  \foreach \i in {ref, exi, desc}
    \draw[flow] (\i.east) -- (llm.west);

  \draw[flow] (llm) --
    node[note, above] {writes} (gen);

  \node[chk, right=9mm of gen] (g2)
    {G2 Robustness};

  \node[chk, above=2.5mm of g2] (g1)
    {G1 Structural Correctness};

  \node[chk, below=2.5mm of g2] (g3)
    {G3 Input Validity};

  \draw[flow] (gen) -- (g2);

  \coordinate (top) at ($(ref.north)+(0,0.5mm)$);
  \coordinate (bot) at ($(desc.south)+(0,-0.5mm)$);

  \begin{scope}[on background layer]
    \node[stage,
      fit=(llm)(gen)(top -| gen)(bot -| gen)] (s1) {};
    \node[stage,
      fit=(g1)(g3)(top -| g2)(bot -| g2)] (s2) {};
  \end{scope}

  \node[title, anchor=south] at (s1.north)
    {\ding{172} Input Generator Synthesis};

  \node[title, anchor=south] at (s2.north)
    {\ding{173} Validation \& Repair};

  \node[title, anchor=south]
    at (ref.north |- s1.north) {Inputs};

  \coordinate (repairbottom) at
    ($(s2.south)+(0,-4mm)$);

  \draw[->, semithick, red!70!black]
    ([xshift=-6mm]s2.south)
    -- ++(0,-4mm)
    -| (llm.south |- s1.south);

    \node[
      note,
      text=red!70!black,
      fill=white,
      inner sep=1pt,
      anchor=west
    ] at ($(s1.south west)+(13mm,-4mm)$)
      {fail: error messages and failing examples};

  \node[box, text width=25mm, below=30mm of g3] (gi)
    {Validated\\[-1pt]input generator};

  \draw[flow] (g3.south) --
    node[note, right, pos=0.65] {pass}
    (gi.north);

  \node[
    box, text width=25mm,
    fill=green!10,
    below=7mm of gi,
    minimum height=7mm
  ] (fz)
    {{\large\faRandom}\ Fuzzer
     {\scriptsize(Atheris/Jazzer)}};

  \draw[flow] (fz) --
    node[note, right] {bytes} (gi);

  \node[prog, left=16mm of gi, yshift=10mm] (pr)
    {Reference $R$};

  \node[prog, left=16mm of gi, yshift=-10mm] (pc)
    {Candidate $C$};

  \begin{scope}[on background layer]
    \node[
      draw=blue!45,
      rounded corners=3pt,
      fill=blue!2,
      inner xsep=3.5mm,
      inner ysep=3mm,
      fit=(pr)(pc)
    ] (programs) {};
  \end{scope}

  \node[
    font=\sffamily\scriptsize,
    align=center,
    fill=blue!2,
    inner sep=1pt
  ] at (programs.center)
    {Combined fuzz target};

  \coordinate (mid) at
    ($(pr.west)!0.5!(pc.west)$);

  \node[
    box,
    text width=15mm,
    left=10mm of mid
  ] (cmp)
    {{\Large\faCogs}\\[1pt]
     Compare\\[-1pt]outputs};

  \draw[flow]
    (gi.west) --
    node[note, above, sloped, pos=0.55] {input}
    (pr.east);

  \draw[flow]
    (gi.west) --
    node[note, below, sloped, pos=0.55] {input}
    (pc.east);

  \draw[flow] (pr.west) -- (cmp.east);
  \draw[flow] (pc.west) -- (cmp.east);

  \node[
    box,
    text width=32mm,
    align=left,
    fill=gray!8,
    left=5mm of cmp
  ] (rep)
    {{\large\faFile*[regular]}\ \textbf{HTML report}\\[2pt]
     \scriptsize$\bullet$ similarity score\\[-1pt]
     \scriptsize$\bullet$ counterexamples\\[-1pt]
     \scriptsize$\bullet$ coverage of $R$ and $C$};

  \draw[flow] (cmp.west) -- (rep.east);


  \coordinate (covstart) at
    ($(programs.south east)+(0mm,1.5mm)$);

  \draw[
    ->,
    semithick,
    dashed
  ]
    (covstart) 
    -- node[note, above, pos=0.55, align=center]
       {coverage}
       node[note, below, pos=0.55, align=center]
       {of $R$, $C$}
    (fz.west);

  \begin{scope}[on background layer]
    \node[
      stage,
      fit=(gi)(fz)(programs)(cmp)
    ] (s3) {};
  \end{scope}

  \node[title, anchor=south]
    at (s3.north)
    {\ding{174} Coverage-guided Differential Fuzzing};

    \node[title, anchor=south]
      at ($(rep.north |- s3.north)+(0,-9mm)$)
      {Output};
\ignore{
  \node[
    inp,
    text width=34mm,
    anchor=north,
    minimum height=7mm
  ] (cand)
    at ($(s3.south -| pc)+(0,-4mm)$)
    {Candidate(s) $C$ {\scriptsize(input)}};

  \draw[flow]
    (cand.north) -- (pc.south);
}
\end{tikzpicture}%
}
\caption{Overview of \textsc{Difuzzi}. The input generator
is synthesized and validated once per reference program
and reused for each candidate.}
\label{fig:workflow}
\end{figure}
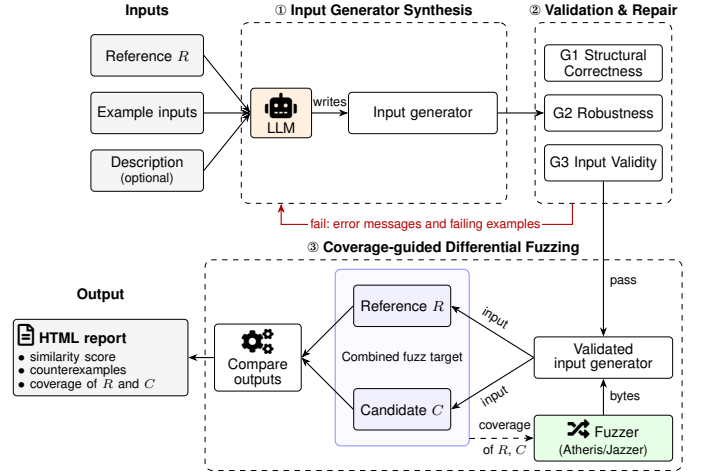

Fig.~\ref{fig:workflow} presents the overall architecture of \textsc{Difuzzi}, highlighting the three main steps discussed below.

\subsubsection{Input Generator Synthesis}
\label{sec: iput_generator}
Since \dfuzz uses fuzzers that operate on raw bytes without knowledge of program-specific input constraints, it is necessary that the test inputs remain valid and relevant to the programs under test. To address this, step 1 uses few-shot prompting to have an LLM synthesize an input generator, given $R$, example inputs, and an optional NL description. The synthesized generator transforms raw fuzzer bytes into valid inputs for $R$.  As shown in Fig.~\ref{fig:example_step1}, the synthesized generator for binary search produces sorted lists of distinct integers and varies the search target $t$ by either selecting an element from the list or by generating an independent integer. This ensures that the generated inputs satisfy $R$'s constraints and cover both scenarios: where the target is present in the list and where it is absent.

\subsubsection{Validation and Repair}
In Step 2, \dfuzz identifies and addresses potential errors in LLM-synthesized input generators before fuzzing. To eliminate manual effort, \dfuzz applies three automated guardrails, each targeting a specific property:

\begin{itemize}
    \item \textbf{G1 (Structural Correctness):}  Verifies that the generator compiles successfully and implements the required \texttt{generate} (See Fig.~\ref{fig:example_step1}) interface.

    \item \textbf{G2 (Robustness):} Executes the input generator on 100 randomly generated byte sequences, checking that each produces an input without exceptions or timeouts.

    \item \textbf{G3 (Input Validity):} Checks whether each input generated during G2 is valid for  $R$. An input is considered valid if $R$ returns or exits normally within the time limit.

\end{itemize}

If any check fails, \dfuzz feeds the error messages and failing examples back to the LLM to guide generator repair. If no generator passes all checks within three attempts, \dfuzz reports the unresolved failures and terminates without estimating functional similarity for the reference and its candidates.

\subsubsection{Coverage-guided Differential Fuzzing}

In the final step, the validated input generator is integrated with a coverage-guided fuzzer. A single fuzz target combining both $R$ and $C$ is constructed, where each generated input is passed to both implementations and their outputs are compared. Note that while the LLM-synthesized generator (Section~\ref{sec: iput_generator}) defines the input space, coverage-guided fuzzing drives the exploration of that space towards diverse inputs. The generator is excluded from coverage instrumentation so that fuzzing is guided solely by coverage from $R$ and $C$. Duplicate inputs are discarded, and fuzzing continues until the target number of distinct inputs (1,000 by default) or the time limit (60\,s by default) is reached.

\textbf{Similarity score.} \dfuzz measures functional similarity as the fraction of evaluated inputs $X$ on which $R$ and $C$ exhibit the same outcome (identical outputs or both failing):
\begin{equation}
  \mathit{sim}(R,C) =
  |\{x \in X : R(x) \simeq C(x)\}| / |X| \in [0,1].
  \label{eq:sim}
\end{equation}

where a score of 1 indicates that no evaluated input distinguishes the two programs, with lower scores reflecting a greater proportion of inputs on which $C$ deviates from $R$. When a binary decision is needed, the score can be mapped to a verdict:
\ignore{
\begin{equation}
  \mathit{eq}(R, C) =
  \begin{cases}
    \textit{equivalent}     & \text{if } \mathit{sim}(R, C) = 1,\\
    \textit{not equivalent} & \text{if } \mathit{sim}(R, C) < 1.
  \end{cases}
  \label{eq:verdict}
\end{equation}
}
$eq(R,C) = (sim(R,C) = 1)$.

Each execution produces an HTML report (See \url{https://github.com/Dristi123/difuzzi_tool}) summarizing functional similarity, identified counterexamples, and line and branch coverage of $R$ and $C$.

\subsection{Implementation and Tool Usage}
\label{sec:usage}

\textsc{Difuzzi} is written in about 2,600 lines of Python code, requiring Python 3.12+ and JDK 17-19 for Java support. For Python, it uses Atheris~\cite{atheris2020} as the fuzzer and
\texttt{coverage.py} for coverage measurement, while Java uses
Jazzer~\cite{jazzer_sp26} and JaCoCo, respectively. \dfuzz supports Kimi and Google Gemini APIs and allows flexible model selection from either provider. A typical invocation for evaluating \textit{cand.py} ($C$) against \textit{ref.py} ($R$) with input examples in \textit{ex.txt} is:
\begin{tcolorbox}[
    colback=gray!5,
    colframe=gray!40,
    boxrule=0.4pt,
    arc=2pt,
    left=4pt,
    right=4pt,
    top=2pt,
    bottom=2pt
]
\begin{Verbatim}[fontsize=\small, fontfamily=tt, fontseries=b]
difuzzi ref.py cand.py --examples ex.txt 
\end{Verbatim}
\end{tcolorbox}

\dfuzz provides 14 command-line options for flexible configuration, including \texttt{--early-stop}, which terminates fuzzing at the first disagreement and returns a counterexample, thereby reducing testing cost when only a binary verdict is needed. All other command-line options, implementation details and setup instructions are documented in the repository.

\section{Evaluation}
\label{sec:evaluation}
\subsection{Dataset} We build our evaluation dataset using five widely used coding benchmarks: CodeContests-Python, HumanEval+, and MBPP+ for Python, along with CodeContests-Java and HumanEval-X for Java, sampling 50 tasks from each. For each task, we pair the reference implementation $R$ with two candidate implementations $C$: a single-edit mutant generated using MutPy~\cite{mutpy} (Python) or Major~\cite{major} (Java), and a non-mutant implementation. While mutants introduce subtle code changes, non-mutants are expected to exhibit greater structural differences from the reference. Non-mutants are drawn from human-written solutions in CodeContests or generated using Qwen 2.5 Coder-32B for the other benchmarks that lack human-written solutions. This yields 500 $R-C$ pairs.

\par To approximate ground-truth functional similarity, we execute $R$ and $C$ on the benchmark-provided test suites and compute their output agreement (Eq.~\ref{eq:sim}). The median number of tests used to estimate ground-truth similarity is 944 for HumanEval+, 108 for MBPP+, 888 for HumanEval-X, and 201 for CodeContests (Python and Java). 

\subsection{Baselines} 
We compare \dfuzz against four baselines that can provide continuous similarity scores: CodeScore~\cite{codescore}, a CEM, and three LLM judges. We select CodeScore because it is trained on program execution outcomes, unlike other commonly used CEMs that rely on textual similarity to reference implementations. For the LLM judges, we use Kimi K2.5, Gemini 2.5 Flash, and Qwen 2.5 Coder-32B.

\subsection{Experimental Setup}

We prompt each LLM-judge to assign a functional similarity score in $[0,1]$ given the task description, the reference and the candidate. For CodeScore, we use its released pretrained checkpoint and treat its predicted pass ratio as the similarity score. Since CodeScore is trained on Python, we evaluate it only on Python pairs. We run \dfuzz with the task description, a single example input, its default budgets, and Kimi K2.5 for input generator synthesis. Table~\ref{tab:similarity} reports Spearman's $\rho$ between each technique's score and the ground-truth.

\begin{table}[t]
\centering
\caption{Spearman's $\rho$ with the ground-truth similarity. The ``All" column is computed over all 500 pairs}
\label{tab:similarity}
\footnotesize
\setlength{\tabcolsep}{3.6pt}
\begin{tabular*}{\columnwidth}{@{\extracolsep{\fill}}lccccc@{}}
\toprule
 & \multicolumn{2}{c}{Python} & \multicolumn{2}{c}{Java} & \\
\cmidrule(lr){2-3}\cmidrule(lr){4-5}
Technique & Mutant & Non-mutant & Mutant & Non-mutant & All \\
\midrule
CodeScore        & 0.31 & 0.52 & -- & -- & 0.35 \\
Kimi K2.5        & 0.51 & 0.77 & 0.84 & 0.84 & 0.86 \\
Gemini 2.5 Flash & 0.65 & 0.57 & 0.80 & 0.62 & 0.81 \\
Qwen 2.5 Coder    & 0.33 & 0.40 & 0.51 & 0.32 & 0.63 \\
\midrule
\dfuzz           & \textbf{0.81} & \textbf{0.90} & \textbf{0.92} & \textbf{0.91} & \textbf{0.94} \\
\bottomrule
\end{tabular*}
\end{table}

\subsection{Results and Analysis} 

\dfuzz correlates most strongly with the ground-truth across all settings (Table~\ref{tab:similarity}), reaching $\rho=0.94$ overall, compared to 0.86 for the best baseline. CodeScore and Qwen 2.5 Coder exhibit weak correlations across all settings, whereas the closed-source models (Kimi K2.5, Gemini 2.5 Flash) achieve comparatively stronger results. Interestingly, no single LLM judge consistently outperforms the others across all settings (on Python mutants, Gemini leads Kimi, $\rho=0.65$ vs.\ 0.51; on Python non-mutants, Kimi leads Gemini, 0.77 vs.\ 0.57). This variation raises concerns about relying on a single LLM judge across settings, in contrast to \dfuzz, which outperforms all baselines throughout.

\par We further compare \dfuzz with the LLM judges on non-equivalence detection and counterexample validity (Table~\ref{tab:detection}). For non-equivalence detection, we measure the percentage of ground-truth non-equivalent pairs that each technique correctly identifies (similarity score $<$ 1, Eq.~\ref{eq:sim}). For pairs identified as non-equivalent, we ask each LLM to provide up to 20 counterexamples, the same maximum number in \dfuzz's report. Counterexample validity is measured as the percentage of reported inputs that produce different outputs when executed on $R$ and $C$.

As Table~\ref{tab:detection} shows, the LLM judges detect nearly all non-equivalent single-edit mutants, but their detection rates drop substantially for non-mutants (Kimi: 82\%, Gemini: 75\%, Qwen: 51\%), compared to 92\% for \dfuzz. We hypothesize that LLMs may have learned common mutation patterns during training, allowing them to identify single-edit mutants easily, whereas the greater structural diversity of independently written human and LLM solutions makes non-equivalence harder to detect. Counterexample validity similarly drops to 86\% for Kimi and 11\% for Qwen on non-mutants. In contrast, \dfuzz achieves 100\% counterexample validity by construction, as every reported counterexample is verified through execution to produce different outputs for $R$ and $C$. Beyond these empirical results, \dfuzz offers a conceptual advantage: its coverage-guided fuzzing and reported coverage statistics allow users to assess how extensively the programs were tested and thereby better interpret the similarity scores, whereas LLM judges lack comparable evidence of testing thoroughness.

\par Finally, we evaluate the quality of the input generators through coverage and input validity. During our experiments, \dfuzz achieved mean branch coverage of 96\% for Python and 99\% for Java references, providing strong evidence of diverse program behavior exploration. Moreover, only 8 of the 459K inputs generated by \dfuzz ($<0.01\%$) caused a reference failure, suggesting that the generators largely produce valid inputs. Together, these results support the effectiveness of our input generator synthesis and validation steps.

\begin{table}[t]
\centering
\caption{Non-equivalence detection and counterexample validity.}
\label{tab:detection}
\footnotesize
\setlength{\tabcolsep}{2.5pt}
\begin{tabular*}{\columnwidth}{@{\extracolsep{\fill}}lcccc@{}}
\toprule
 & \multicolumn{2}{c}{\shortstack{Non-equivalence (\%)}}
 & \multicolumn{2}{c}{\shortstack{Cex. Validity (\%)}} \\
\cmidrule(lr){2-3}\cmidrule(l){4-5}
Technique & Mutant & Non-mut. & Mutant & Non-mut. \\
\midrule
Kimi K2.5        & \textbf{100} & 82 & 93 & 86 \\
Gemini 2.5 Flash & \textbf{100} & 75 & 90 & 72 \\
Qwen 2.5 Coder    & 95 & 51 & 43 & 11 \\
\midrule
\dfuzz           & \textbf{100} & \textbf{92} & \textbf{100} & \textbf{100} \\
\bottomrule
\end{tabular*}
\vspace{-0.2mm}
\end{table}

\section{Applications}

The differential fuzzing approach underlying \dfuzz has supported three of our prior studies addressing distinct software engineering problems. In~\cite{Local}, it was used to build a benchmark with ground-truth functional similarity scores for evaluating CEMs. In our study of LLM-generated code refactorings~\cite{diffuzz_refac}, we applied it to assess functional equivalence between original and refactored code, uncovering behavioral differences missed by existing test suites. In ~\cite{dristi2026sonartaskawarecodesummary}, differential fuzzing supported code summary evaluation by measuring the functional similarity between the original code and code regenerated from summaries. However, these studies required substantial manual effort to configure and execute differential fuzzing, motivating us to refine and automate the workflow and make \dfuzz publicly available. We envision \dfuzz's broader applications in regression testing, validation of automated program repairs, code optimizations, and other tasks requiring functional similarity estimates, which \dfuzz can provide without relying on pre-existing test cases.

\section{Threats to Validity}
The validity and input diversity of LLM-synthesized generators may affect our findings. To mitigate this risk, we validate and repair generators before fuzzing and rely on coverage-guided fuzzing, rather than the LLM, to explore the input space. The ground-truth similarity estimates (Section~\ref{sec:evaluation}) may not reflect true functional similarity, as they are derived from finite test suites. Nevertheless, we use extended test suites where available to obtain a reasonable approximation of ground truth. Finally, although \dfuzz generates a large number of inputs per pair, we note that agreement on these inputs is not a formal proof of functional equivalence.

\section{Conclusion}
This work presents \dfuzz, an open-source differential fuzzing tool for estimating functional similarity without pre-existing test suites. It  uses LLM-based input generator synthesis, automated validation, and coverage-guided fuzzing to estimate functional similarity and identify counterexamples. Experiments demonstrate that \dfuzz achieves very strong correlations with ground-truth scores, while ensuring 100\% counterexample validity by construction. Moreover, the synthesized input generators achieve high coverage and input validity, demonstrating that \dfuzz can reliably automate the workflow. We hope that the provided repository, documentation, and demonstration video will facilitate broader adoption of \dfuzz. As future work, we plan to extend \dfuzz to support cross-language functional similarity assessment and broaden its scope of observable behavior beyond return values and standard output to include side effects such as changes to global state and file writes.

\bibliographystyle{IEEEtran}
\bibliography{ref}

@IEEEtranBSTCTL{IEEEexample:BSTcontrol,
  CTLuse_forced_etal       = "yes",
  CTLmax_names_forced_etal = "3",
  CTLnames_show_etal       = "1",
  CTLuse_url               = "no"
}

@inproceedings{diffuzz_refac,
author = {Dristi, Simantika Bhattacharjee and Dwyer, Matthew B.},
title = {A Differential Fuzzing-Based Evaluation of Functional Equivalence in LLM-Generated Code Refactorings},
year = {2026},
isbn = {9798400726361},
booktitle = {Proceedings of the 34th ACM International Conference on the Foundations of Software Engineering},
pages = {1708–1712},
numpages = {5},
location = {Montreal, QC, Canada},
series = {FSE Companion '26}
}

@article{Local,
author = {Dristi, Simantika Bhattacharjee and Dwyer, Matthew B.},
title = {LoCaL: Countering Surface Bias in Code Evaluation Metrics},
year = {2026},
issue_date = {July 2026},
publisher = {Association for Computing Machinery},
address = {New York, NY, USA},
volume = {3},
number = {FSE},
journal = {Proc. ACM Softw. Eng.},
month = jun,
articleno = {FSE061},
numpages = {24}
}

@inproceedings{
shypula2024learning,
title={Learning Performance-Improving Code Edits},
author={Alexander G Shypula and Aman Madaan and Yimeng Zeng and Uri Alon and Jacob R. Gardner and Yiming Yang and Milad Hashemi and Graham Neubig and Parthasarathy Ranganathan and Osbert Bastani and Amir Yazdanbakhsh},
booktitle={The Twelfth International Conference on Learning Representations},
year={2024},
}

@misc{atheris2020,
  title        = {{Atheris}: A Coverage-Guided, Native Python Fuzzer},
  howpublished = {\url{https://github.com/google/atheris}},
  year         = {2020},
}

@inproceedings{jazzer_sp26,
  author={Dechand, Sergej and Wienand, Tobias and Meumertzheim, Fabian and Samarin, Peter and Resch, Simon and Yakdan, Khaled and Holz, Thorsten and Toffalini, Flavio},
  booktitle={2026 IEEE Symposium on Security and Privacy (SP)},
  title={Jazzer: Coverage-Guided Fuzzing for Semantic Vulnerabilities in the Java Ecosystem},
  year={2026},
  pages={3720-3739},
}

@misc{ren2020codebleumethodautomaticevaluation,
      title={CodeBLEU: a Method for Automatic Evaluation of Code Synthesis}, 
      author={Shuo Ren and Daya Guo and Shuai Lu and Long Zhou and Shujie Liu and Duyu Tang and Neel Sundaresan and Ming Zhou and Ambrosio Blanco and Shuai Ma},
      year={2020},
      eprint={2009.10297},
      archivePrefix={arXiv},
      primaryClass={cs.SE},
      url={https://arxiv.org/abs/2009.10297}, 
}

@inproceedings{equibench,
    title = "{E}qui{B}ench: Benchmarking Large Language Models' Reasoning about Program Semantics via Equivalence Checking",
    author = "Wei, Anjiang  and
      Cao, Jiannan  and
      Li, Ran  and
      Chen, Hongyu  and
      Zhang, Yuhui  and
      Wang, Ziheng  and
      Liu, Yuan  and
      Teixeira, Thiago S. F. X.  and
      Yang, Diyi  and
      Wang, Ke  and
      Aiken, Alex",
    booktitle = "Proceedings of the 2025 Conference on Empirical Methods in Natural Language Processing",
    month = nov,
    year = "2025",
    pages = "33868--33881",
    ISBN = "979-8-89176-332-6"
}

@misc{crosshair,
  author       = {Philip Schanely},
  title        = {CrossHair},
  howpublished = {\url{https://github.com/pschanely/CrossHair}},
  note         = {GitHub repository, accessed October 19, 2026},
  year         = {2026}
}

@article{mokav,
   title={Mokav: Execution-driven differential testing with LLMs},
   volume={230},
   ISSN={0164-1212},
   journal={Journal of Systems and Software},
   publisher={Elsevier BV},
   author={Etemadi, Khashayar and Mohammadi, Bardia and Su, Zhendong and Monperrus, Martin},
   year={2025},
   month=Dec, pages={112571} }

@article{codescore,
author = {Dong, Yihong and Ding, Jiazheng and Jiang, Xue and Li, Ge and Li, Zhuo and Jin, Zhi},
title = {CodeScore: Evaluating Code Generation by Learning Code Execution},
year = {2025},
issue_date = {March 2025},
volume = {34},
number = {3},
issn = {1049-331X},
journal = {ACM Trans. Softw. Eng. Methodol.},
month = feb,
articleno = {77},
numpages = {22}
}

@misc{meeus2026detectingfunctionalmemorizationcode,
      title={Detecting Functional Memorization in Code Language Models}, 
      author={Matthieu Meeus and Anil Ramakrishna and Shengyuan Hu and Matthew Grange and Zheng Xu and Luca Melis},
      year={2026},
      eprint={2606.12764},
      archivePrefix={arXiv},
      primaryClass={cs.LG},
      url={https://arxiv.org/abs/2606.12764}, 
}

@misc{dristi2026sonartaskawarecodesummary,
      title={SONAR: Task-Aware Code Summary Evaluation for LLM Consumers Without References}, 
      author={Simantika Bhattacharjee Dristi and Matthew B. Dwyer},
      year={2026},
      eprint={2608.04195},
      archivePrefix={arXiv},
      primaryClass={cs.SE},
      url={https://arxiv.org/abs/2608.04195}, 
}

@misc{mutpy,
  author = {Konrad Ha{\l}as and Contributors},
  title = {{MutPy}: Mutation Testing Tool for Python},
  year = {2015},
  howpublished = {\url{https://github.com/mutpy/mutpy}}
}

@inproceedings{major,
  title={The major mutation framework: efficient and scalable mutation analysis for Java},
  author={Ren{\'e} Just},
  booktitle={International Symposium on Software Testing and Analysis},
  year={2014},
}

@misc{difuzzi,
  title        = {{DIFUZZI}: A Differential Fuzzing Tool for Code Functional Similarity Assessment},
  author       = {Simantika Bhattacharjee Dristi and Matthew B. Dwyer},
  howpublished = {\url{https://github.com/Dristi123/difuzzi}},
  year         = {2026}
}

@misc{debug,
      title={Benchmarking Code Improvement with Progressive, Adaptive, and Interactive Feedback}, 
      author={Cuong Chi Le and Aashish Yadavally and Minh Le-Anh and Tien N. Nguyen},
      year={2026},
      eprint={2607.01360},
      archivePrefix={arXiv},
      primaryClass={cs.SE},
      url={https://arxiv.org/abs/2607.01360}, 
}

@INPROCEEDINGS{10795375,
  author={Siddiq, Mohammed Latif and Dristi, Simantika and Saha, Joy and Santos, Joanna C. S.},
  booktitle={2024 IEEE International Conference on Source Code Analysis and Manipulation (SCAM)}, 
  title={The Fault in our Stars: Quality Assessment of Code Generation Benchmarks}, 
  year={2024},
  volume={},
  number={},
  pages={201-212},
  doi={10.1109/SCAM63643.2024.00028}}
\end{document}